**Suppression of compensating native defect formation during semiconductor processing via excess carriers**

K. Alberi[1]* and M.A. Scarpulla[2,3]

[1.] *National Renewable Energy Laboratory, Golden CO 80401*

[2.] *Material Science and Engineering, University of Utah, Salt Lake City, UT 84112*

[3.] *Electrical and Computer Engineering, University of Utah, Salt Lake City, UT 84112*

In many semiconductors, compensating defects set doping limits, decrease carrier mobility, and reduce minority carrier lifetime thus limiting their utility in devices. Native defects are responsible in many cases, but extrinsic dopants may also act as their own compensation when occupying an alternate lattice site. Suppressing the concentrations of compensating defects during processing close to thermal equilibrium is difficult because formation enthalpies are lowered as the Fermi level moves towards the majority band edge. Excess carriers, introduced for example by photogeneration, modify the formation enthalpy of semiconductor defects and thus can be harnessed during crystal growth or annealing to suppress such defect populations. Herein we develop a rigorous and general model for defect formation in the presence of steady-state excess carrier concentrations by combining the standard quasi-chemical formalism with a detailed-balance description that is applicable for any defect state in the bandgap. Considering the quasi-Fermi levels as chemical potentials, we demonstrate that increasing the minority carrier concentration increases the formation enthalpy for typical compensating centers, thus suppressing their formation. This effect is illustrated for the specific example of native $Ga_{Sb}$ antisite acceptors in nominally un-doped and extrinsically n-type doped GaSb. The model also predicts reductions in the concentration of ionized scattering centers as well as increased mobility. While our treatment is generalized for excess carrier injection or generation in semiconductors by any means, we provide a set of guidelines for applying the concept in photoassisted physical vapor deposition.

Compensating native defects degrade semiconductor properties and device performance by introducing effects such as doping limits, trapping, and defect-assisted recombination. Modern uses of semiconductors therefore critically depend in large part on eliminating such centers. While both extrinsic impurities and native defects can introduce states in the bandgap, the latter are the root cause of particularly tenacious materials challenges for two reasons. First, they can be generated in arbitrarily-high numbers from the perfect crystal. Second, the free energies of formation for their non-zero charge states are dependent on the Fermi level and their charge transition levels. For instance, adding an extrinsic n-type dopant raises the Fermi level, which in turn reduces the formation energy for compensating, ionized acceptor-like native defects. The result is a limit on the achievable free carrier concentration that has been described by the concept of a Fermi stabilization energy.[1,2] This negative feedback inherent to defect equilibrium forms the basis of many decades-old material technology challenges that severely hinder or prevent the use of certain semiconductors with otherwise superior properties.

Overcoming the challenges caused by native defects has been particularly difficult because their equilibrium concentration in the bulk and at interfaces is controlled by only two types of tunable thermodynamic potentials: 1) temperature and 2) chemical potentials. This assumes that the electron and hole populations, the lattice, and the electromagnetic field are all in (at least local) thermal equilibrium. However, the application of non-equilibrium conditions can provide access to additional independent and tunable process variables. Excess photogenerated carrier populations are an especially promising choice for controlling defect densities because they directly affect the quasi-Fermi levels (QFLs). Since the Fermi level governs particle and energy exchange between the defect and the free carrier reservoirs, splitting the QFLs should be able to enhance or suppress defect formation by altering this exchange. The use of light to affect semiconductor growth has been tentatively studied[3,4,5], and there is some evidence linking doping enhancement to defect reduction. Yet, the mechanisms underpinning the improvements were not well developed. Understanding the relationships and interactions between defects, photogenerated carriers and the Fermi level is critical for assessing the potential for defect suppression via light stimulated processing techniques and can ultimately direct their implementation.

Herein, we develop a complete framework for determining the interplay of steady-state excess carrier generation, trapping and recombination, and native defect formation to allow prediction of native defect populations during semiconductor processing. Our theoretical framework self-consistently determines the free carrier concentrations, quasi-Fermi levels, recombination rates, defect occupations, and defect concentrations. In order to illustrate its utility, we apply this model to the concrete example of compensating native acceptor suppression during light-assisted epitaxy or annealing of n-type GaSb. This material system exhibits dominant native defects of either positive or negative charge depending on the prevailing chemical potentials during processing and is therefore ideal for assessing the influence of each. From these results, we have constructed a set of guidelines that can be applied to the light-assisted processing of semiconductors. Our findings have implications for physical vapor deposition techniques such as evaporation, hydride vapor phase epitaxy and molecular beam epitaxy (MBE) and also especially for understanding differences between furnace and lamp-heated annealing treatments such as rapid thermal processing (RTP).

**Results**

**Theoretical Framework**

At thermal and chemical equilibrium, the concentrations of point defects within a material phase describe a unique thermodynamic state and are found in general by minimizing the appropriate free energy. It is important to remember that this thermodynamic equilibrium is microscopically a state of dynamic equilibrium. Individual point defects are created and annihilated at certain rates resulting in steady-state concentrations with negligible fluctuations for large system sizes. When the dominant entropy is configurational in the dilute limit, the system can be treated within the quasi-chemical formalism[6,7]. The concentrations of N species of point defects can be computed if the enthalpies of formation or quasi-chemical reaction constants are measured or calculated. A system of N+2 equations, consisting of N quasi-chemical defect creation reactions, the electron/hole pair generation reaction (or law of mass-action $np=n_i^2$), and the charge balance equation can be solved at a given temperature by self-consistently determining the Fermi level. In particular, the dependence of the defect formation energy on the Fermi level is introduced through energy transfer between the charge transition level of the defect and the appropriate reservoir of hole or electron carriers (i.e. the valence or conduction band). The chemical potential that characterizes the reservoir is simply the Fermi level when the bands of the semiconductor are in thermal equilibrium.

Under conditions of steady-state photocarrier generation, we must further consider the variation of defects' formation energies when the electron and hole QFLs split. It is well established (e.g. in analysis of open circuit voltage of solar cells) that the QFLs represent the electrochemical potentials of the electron and hole populations in the conduction and valence bands of the crystal once those populations have thermalized amongst themselves and with the lattice but prior to recombination.[8] Defects may exist in multiple charge states and may generally capture both electrons and holes via the electron-phonon interaction, as described by their capture cross-sections.[9] The occupancy of defect charge states depends on both the free electron and hole concentrations. Consequently, the formation free energies for all defects in the system are a function of both QFLs, weighted by the strength or frequency of interactions with the valence and conduction bands.

Because charged states arise through real forward and reverse exchange of carriers between the localized defect and the conduction and valence bands, the most accurate method of the weighting the dependence of the formation energy on the QFLs that govern this exchange is as a ratio of the rates of these transactions. To illustrate this point, consider the formation of a defect of charge state $q$ = -1. This state can be produced by either the capture of an electron or the emission of a hole. The electron and hole QFLs would then be weighted based on these two exchange rates, respectively.

A consequence of this behavior is that the defect formation energy will be tied most strongly to the QFL that governs the dominant exchange rate. For instance, negatively charged defects are typically more likely to trap holes, and thus their concentration will be affected more by the energy of the hole QFL than the electron QFL. However, it is important to bear in mind that the defect's interaction with both bands and corresponding dependence on both QFLs is in stark contrast to a shallow dopant, which exchanges carriers exclusively with only one band. The full details of the theoretical approach are included in the Computational Methods section.

**Defect Calculations**

To understand the potential effects of illumination on defect concentrations during light-assisted semiconductor processing, such as epitaxy or rapid thermal annealing, we have applied this framework to the model system of n-type GaSb. GaSb is an interesting but simple system in which to study the effects of light-based processing because of its propensity to form a single species of compensating defect over the full range of chemical potentials. Furthermore, excessive native defect concentrations and apparent doping limits impact its performance in practical applications.

Epitaxial growth is typically carried out in the temperature range 400-550 °C. Under Ga-rich growth conditions, negatively charged $Ga_{Sb}$ antisites form with relatively low energy (< 1.8 eV) and are the single most dominant defect in intentionally n-type doped single crystals.[10,11,12] Positively charged $Sb_{Ga}$ antisites and $Ga_i$ interstitials also form at non-negligible concentrations in n-type GaSb grown under Sb-rich conditions.[11] Overall, the formation energies of negatively charged $Ga_{Sb}$ antisites are higher in Sb-rich vs Ga-rich conditions, suggesting that maintaining an Sb-rich environment during GaSb crystal growth or annealing is advantageous for controlling compensating native defect concentrations.

In practice, achieving Sb-rich growth conditions by molecular beam epitaxy (MBE) can be somewhat difficult. Since Sb has a relatively low vapor pressure, the thermally evaporated molecules must be thermally cracked from its initial $Sb_4$ state, and deposited atoms tend to clump on the growth surface, leaving exposed Ga sites. Typical growths are carried out at modest V/III ratios (~2-6) close to those needed to achieve a Sb-stabilized surface. This realistically places the growth in an intermediate regime.[13] Unintentionally doped bulk crystals and epilayers are usually p-type due to high levels of $Ga_{Sb}$ antisites. The presence of charged defect species also degrade the carrier mobility. GaSb has a relatively low bandgap (0.72 eV at room temperature), and the theoretical electron mobility can be as high as 44,000 $cm^{-1}$ V s.[14] However, much lower mobilities are typically observed in part because of scattering by ionized centers.[15] Suppressing defect formation is therefore critical to realizing the theoretical mobilities of GaSb and enabling its widespread use.

**Table 1** Formation energies and ionization levels for the native defects considered in this study. Values from Ref. 11 calculated at 0 K assuming the valence band edge as the zero of energy.

| Defect | Defect Formation Energies (eV) | | Ionization Level (eV) |
|---|---|---|---|
| | *Ga-Rich* | *Sb-Rich* | |
| $Ga_{Sb}^{-1}$ | 1.50 | 2.44 | 0.15 |
| $Ga_{Sb}^{-2}$ | 1.77 | 2.71 | 0.26 |
| $Sb_{Ga}^{+1}$ | 1.91 | 0.97 | Above CB Edge |
| $Sb_{Ga}^{+2}$ | 1.56 | 0.63 | 0.36 |
| $Ga_i^{+1}$ ($T_a$) | 1.38 | 1.85 | Above CB Edge |
| $Ga_i^{+1}$ ($T_c$) | 1.04 | 1.50 | Above CB Edge |

Under all conditions, native defect concentrations rise exponentially with growth temperature, and the total concentrations decrease exponentially with defect formation free energy. The magnitude of any light-induced changes to the defect formation free energy will be bounded (approximately, in degenerate injection conditions) by the bandgap, similar to limits of open circuit voltage in photovoltaics. Likewise, the magnitude of any atomic chemical potential-induced changes to the defect formation free energy are limited by the formation free energy of the compound, which is typically a few eV per formula unit and in general correlates with bandgap. Variation of the photogeneration rate and the chemical potentials are therefore theoretically expected to lead to similar changes in the native defect concentrations. Practically, though, the effects of illumination on native defect populations may be more muted because of the large photon or other particle fluxes required to maintain large QFL splitting, especially in the presence of recombination-active centers and at high temperatures.

To evaluate the relative influence of photogenerated carriers and chemical potentials on the native defect concentrations, calculations were performed as a function of temperature for GaSb crystals under both Ga-rich and Sb-rich environments. The trap species, their ionization levels, and their formation enthalpies used in these calculations are listed in Table 1.[11] Two donor doping concentrations were considered: undoped and moderately doped ($N_d = 10^{17}$ cm$^{-3}$). Additionally, three photon fluxes were applied: 0, $10^{23}$ and $5 \times 10^{24}$ cm$^{-3}$s$^{-1}$. Assuming a carrier lifetime on the order of 1 ns, these rates are expected to produce excess carrier concentrations in the range of $10^{14}$ to $5 \times 10^{15}$ cm$^{-3}$, which is well below the extrinsic n-type doping concentration and are achievable with common lamp and laser light sources.

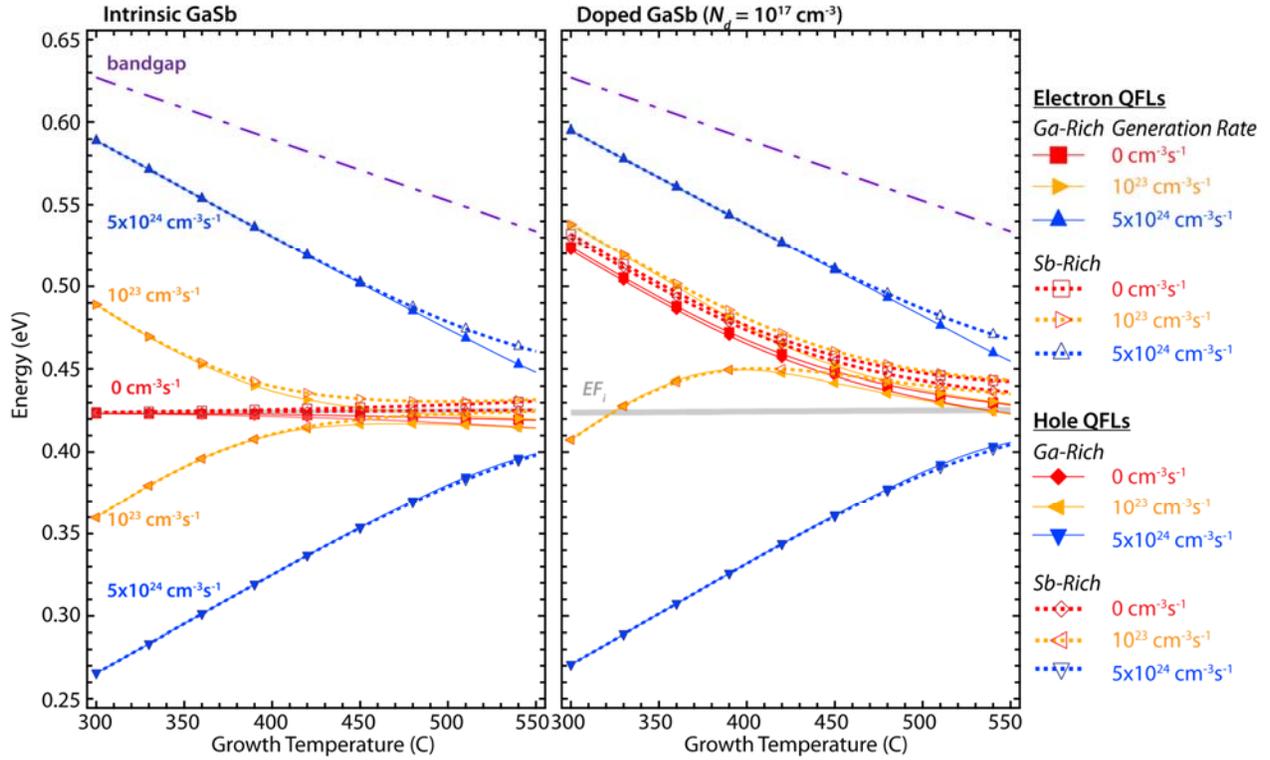

**Figure 1 Fermi levels as a function of growth temperature**. Bandgap (dash-dot), intrinsic Fermi level ($E_{Fi}$ = heavy solid) and electron and hole QFL values calculated for undoped and doped GaSb under Ga-rich (thin solid) and Sb-rich (dashed) growth conditions as functions of temperature and photocarrier generation rate.

Figure 1 shows the calculated QFLs as functions of temperature and excess carrier generation rates for both doping cases. As expected, QFL splitting increases with increasing photocarrier generation rate. The splitting also decreases with increasing growth temperature as the intrinsic carrier concentrations begin to dominate the photo-injected electron and hole concentrations and recombination becomes faster.

The resulting general trends in the concentrations of typical native defects in the presence of excess carriers are determined primarily by a combination of their formation energies, their charge states and the degree of QFL splitting. Our first main finding is that excess carriers always reduce the concentration of charged native defects of both polarities (regardless of extrinsic doping) and that the degree of suppression increases with QFL splitting (Fig. 1). The behaviors of $Ga_{Sb}^{-2}$ and $Sb_{Ga}^{+2}$ antisites exemplify the main trends observed in all of the defects listed in Table 1 and are displayed in Fig. 2. Under Ga-rich growth conditions, $Ga_{Sb}$ antisites are the dominant native defect. Their concentration rises with temperature in undoped GaSb grown under dark conditions as the electron QFL moves closer to the conduction band edge (see Fig. 1) and eventually exceeds $10^{16}$ cm$^{-3}$. These defect concentrations calculated for dark growth conditions are comparable with the excess free hole concentrations measured in nominally un-doped GaSb crystals, which was linked to the presence of native acceptor defects.[10, 16] The higher free electron concentration in the doped case pushes the QFLs closer to the conduction band edge, driving up the $Ga_{Sb}^{-2}$ concentration at all temperatures to approach that of the donor concentration. Compared to these baseline equilibrium defect concentrations in the dark (generation rate = 0 cm$^{-3}$ s$^{-1}$), the suppression under steady-state excitation is most significant at the lowest temperatures due to the larger QFL splitting. In fact, the suppression mechanism is so dominant that the $Ga_{Sb}^{-2}$ concentrations are approximately equal between the undoped and doped cases at the highest generation rate. Under Sb-rich growth conditions, the formation energies of $Sb_{Ga}$ antisites are lower than $Ga_{Sb}$ antisites, making them the dominant defect.

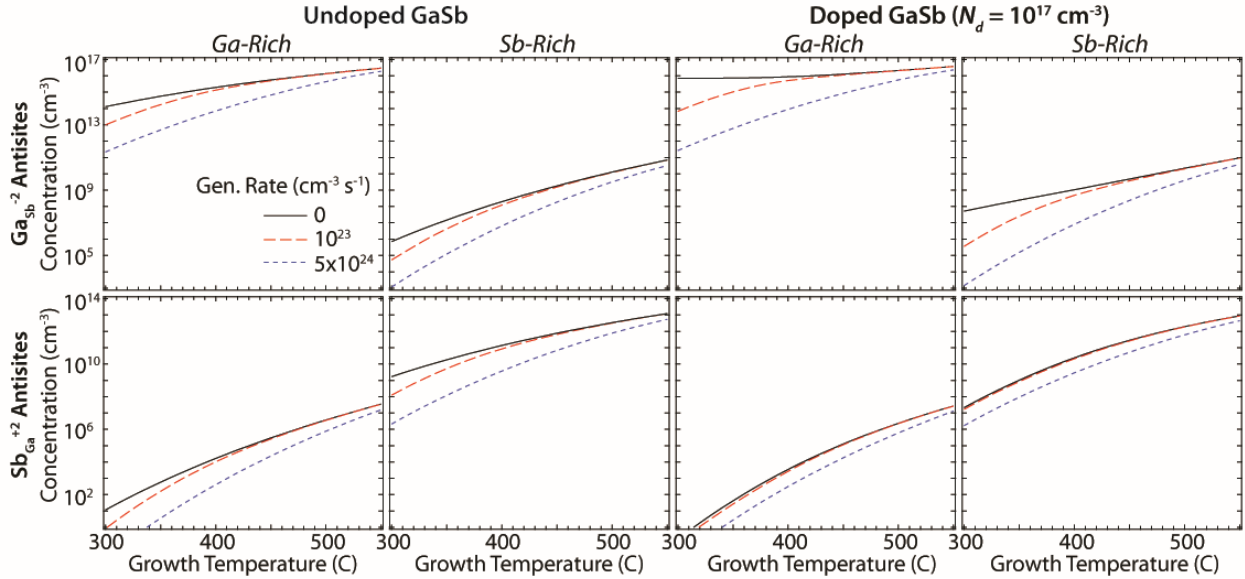

**Figure 2 Defect concentrations as a function of growth temperature.** $Ga_{Sb}^{-2}$ and $Sb_{Ga}^{+2}$ defect concentrations calculated as a function of growth temperature and photogeneration rate for Ga-rich and Sb-rich growth conditions in undoped and doped GaSb.

It is interesting and instructive to compare the behavior of defects with different charge types. For the n-type extrinsic doping case considered here, the presence of photogenerated carriers induce a strong reduction in the compensating $Ga_{Sb}^{-2}$ antisite concentration. The suppression, which is predicted to be approximately 4 orders of magnitude at 300 °C, is most significant in doped GaSb given the higher $Ga_{Sb}^{-2}$ concentration induced by the high Fermi-level energy under dark growth conditions. The suppresion in the undoped case of approximately 3 orders of magnitude is also quite substantial and is only tempered by the lower concentration found under dark conditions. Thus in all cases, photogenerated carriers can be used to substantially reduce the formation of the otherwise compensating native defects.

$Sb_{Ga}^{+2}$ antisites, which have the same charge state as the ionized extrinsic donors, exhibit a more muted change in their concentration in the presence of photogenerated carriers. Additionally, the degree of reduction *decreases* rather than *increases* as the extrinsic shallow donor concentration is increased. The disparity in the response to the presence of photogenerated carriers between these two defects is largely due to their different dependences on the electron and hole QFLs. The $Ga_{Sb}^{-2}$ antisite formation energy is primarily sensitive to the hole QFL, which is significantly affected by the presence of a non-equilibrium concentration of holes at both donor doping levels. On the other hand, the formation energy of $Sb_{Ga}^{+2}$ antisites is strongly tied to the electron QFL. Its energy is altered by the photocarrier generation in undoped GaSb due to the ability to achieve conditions of high injection with relatively low light intensity. In contrast, achieving high injection conditions requires much greater generation rates in more heavily-doped semiconductors. Photocarrier-induced suppression of $Sb_{Ga}^{+2}$ antisites is therefore minimal for the generation rates considered here. Inverse conclusions can be drawn for p-type doped material.

Interestingly, there is a divergence in the QFLs between the Ga-rich and Sb-rich conditions at elevated temperatures. This divergence arises from a difference in the dominant native defect type and concentration, which because of their different charge states and cross sections alter the free electron and hole concentrations. The much higher concentrations of $Ga_{Sb}^{-1}$ and $Ga_{Sb}^{-2}$ defects (~$10^{15}$ – $10^{16}$ cm$^{-3}$) generated under Ga-rich conditions effectively act as additional acceptors and push the QFLs toward the valence band. This result is consistent with the p-type behavior observed in undoped and lightly doped GaSb crystals. The degree of QFL splitting at the same doping level and temperature is also smaller in an Ga-rich vs Sb-rich environment. We attribute this to excessive carrier trapping at the higher concentration of defects generated under those conditions, which effectively regulates the excess carrier population. This result highlights the need to properly account for carrier trapping at defects when calculating the QFL splitting.

In the preceding sections, we have discussed the effects of excess carriers on native defect concentrations and thus dopant compensation and recombination rates. Additional improvements to optoelectronic performance are also realized in light-assisted processing by the higher mobilities made possible by lower compensation and ionized center concentrations. To demonstrate these effects, we calculated the total concentration of all ionized centers (native defects plus donor and acceptor impurities) for undoped and n-type doped GaSb processed under Ga-rich conditions and generation rates of 0 and $5\times10^{24}$ cm$^{-2}$s$^{-1}$. The calculation assumes perfect quenching of the total native defect populations from the growth or processing temperature, and the occupations of the charge states of each defect were recalculated at room temperature.[17] Figure 3a displays these concentrations as a function of growth temperature. The native defect concentration entirely dictates the number of ionized centers in undoped GaSb, and thus the suppression of defect formation by excess carriers leads to a corresponding decrease in the center concentration. A much smaller relative change is observed for n-type doped GaSb due to the additional presence of ionized extrinsic dopants. The extrinsic dopant concentration is of course independent of the presence of photogenerated carriers, and the doping concentration considered here ($10^{17}$ cm$^{-3}$) is high enough to overwhelm the defect concentration. For ionized impurity scattering, the mobility scales inversely with the concentration of each defect multiplied by the square of their charge state. Figure 3b displays the expected mobility ratios resulting from a reduction in the defect concentration under a photocarrier generation rate of $5\times10^{24}$ cm$^{-3}$s$^{-1}$ (compared to the dark case). The upper limit of mobility enhancement is set by the undoped case; for this generation rate and the parameters in Table 1 the ionized impurity mechanism can be suppressed by a factor of ~250 for a growth temperature around 300 °C. This ratio does not include any ionized centers due to shallow dopants, which will ultimately be present at some non-zero concentration in real samples. With increased doping, the mobility enhancement effect decreases, and the processing temperature at which the maximum occurs increases.

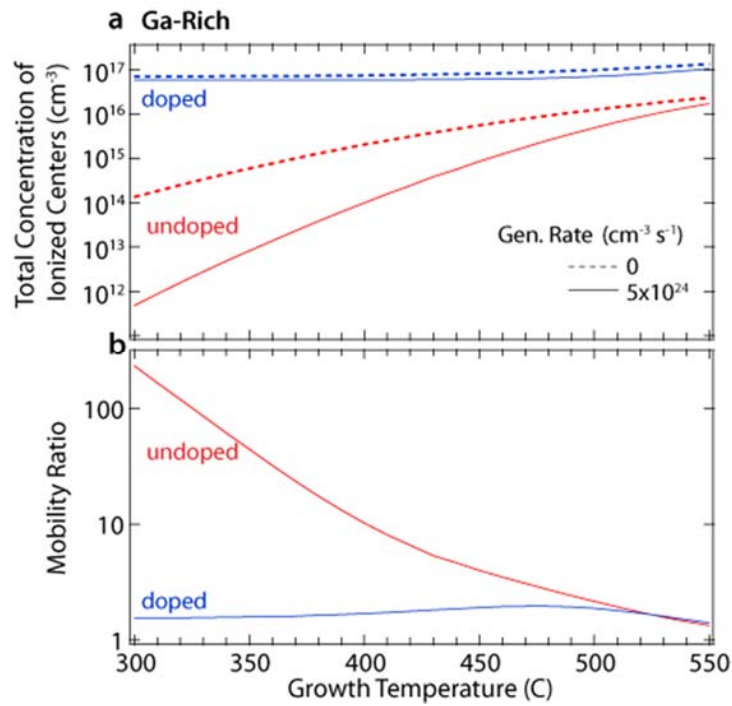

**Figure 3 Room temperature ionized defect data. a,** Total concentration of centers ionized at room temperature in undoped and doped GaSb grown under Ga-rich conditions as a function growth temperature and photogeneration rate. **b,** Ratios of light to dark ionized center-limited mobility from **a**.

**Discussion**

When contemplating the implications and potential for processing semiconductors under a light bias, one should first consider that these calculations apply only to the case of thermodynamic equilibrium. Realization of local equilibrium conditions depends on several factors, including the processing temperature, diffusion coefficients for atoms at the crystal surface and in the bulk, and chemical potentials. Each material will behave differently. However, local equilibrium can be achieved for conventional growth techniques such as molecular beam epitaxy and hydride vapor phase epitaxy, as well as for thermal annealing, with the correct selection of temperatures and time scales (i.e. growth rate, soak time, etc.).

Defect reduction and a general improvement in crystalline quality have been experimentally observed in variety of semiconductors (i.e. CdTe, ZnSe, ZnS, GaN) grown under light stimulation by both molecular beam epitaxy and metal organic chemical vapor deposition.[3,5,18,19,20,21,22] The influence of excess photogenerated carrier populations on defect formation was first explored separately by Marfaing[23] and Ichimura, *et al.*[24] through simplified microscopic models of specific growth processes. More recently, a theoretical model proposed by Bryan, *et al.*, explicitly considered the role of quasi Fermi level splitting on defect populations.[5] The concepts therein represent an advance in defining the intriguing mechanism coupling excess carrier populations to defect concentrations treating quasi-Fermi levels as chemical potentials. However, a complete, rigorous framework was not developed, nor were other mechanisms eliminated as possible explanations for their experimental observations.

Our model offers a comprehensive and accurate description of this specific mechanism of suppressing charged defect formation by altering the electron and hole chemical potentials via injection or generation of excess carriers, which couple to the defect formation enthalpy through capture and emission rates. It defines the conjugate variables to the quasi-Fermi levels (which are the chemical potentials for the free carriers in the bands) as the normalized capture and emission rates between the defect's energy level and appropriate bands (see Eqs. 11a and 11b). This allows us to write the thermodynamic potential $\Delta G = \Delta H - T\Delta S$ (with $\Delta H$ from Eq. 3) which must be minimized in order to find the steady-state concentrations of defects in the light. The quasichemical formalism utilized herein is equivalent to assuming the dilute limit where $\Delta S$ is dominated by configurational entropy.[25]

It is important to emphasize that light or other methods of introducing excess carriers, such as radiation or electron beams[26], may modify doping and compensation via many mechanisms. These include exciting or decomposing precursor molecules, modifying desorption rates[27], reducing surface band bending[28], or creating or annihilating defect complexes. We note the relevance of related bodies of work on photo-induced defects in semiconductors and insulators, for example, color-center formation in II-VI crystals, the Staebler-Wronski effect in amorphous Si, and radiation-enhanced diffusion in GaAs.[29,30,31,32,33,34,35] Carefully-controlled and designed experiments measurements are required to differentiate between the various physical mechanisms. Focused experimental measurements should be made to determine whether enhancements in crystal quality and doping efficiency can be definitively assigned to the mechanism proposed here. Experiments comparing growth to annealing may be particularly helpful in differentiating bulk thermodynamic mechanisms such as the present one from kinetic ones of the growth interface.

The observations presented above naturally lead to a set of common guidelines for controlling native defect concentrations using light-assisted epitaxy or other processing methods during which excess carriers are generated. 1) A greater reduction in the concentration of native defects can be achieved with excess carriers at lower temperatures where the photogenerated carrier population is large enough to support substantial QFL splitting. It is subsequently suppressed when the intrinsic carrier concentration approaches the excess carrier density. From a thermodynamic standpoint, lower temperatures can be beneficial for controlling native defect populations and should be used when it is possible to also suppress defect formation due to kinetic limitations. 2) Increased photogeneration rates result in a greater reduction in native defect concentrations. Practically, there will be an upper limit to the light intensity and density of photogenerated carriers that can be introduced without initiating other processes including: melting, desorption, and crystal damage. There will therefore be an optimal light intensity for each material based on its melting temperature, extrinsic doping level and the vapor pressure of its constituent atoms. 3) The largest changes in concentration are realized for defects that efficiently exchange carriers primarily with the minority carrier reservoir. In typical cases, these will be compensating native defects having shallower charge transition levels and the opposite charge polarity to the extrinsic dopant (and thus larger capture cross section for the minority carrier). In fact, techniques that target the injection of minority carriers (i.e. electrical injection) could also be very effective at reducing compensating defects. Very small effects will be expected in heavily-doped cases for defects of the same charge type as the dominant dopant. 4) In intrinsic or unintentionally-doped cases, the formation of defects of both charge polarities will be more strongly suppressed since both QFLs will change significantly with the introduction of excess carriers. This is also true for recombination at defects interacting equally with both carrier types. 5) The magnitude of compensating defect reduction using this technique will be greater for higher doping concentrations. This is because of the greater driving force for compensating defect formation under dark conditions in the first place. Ultimately, the defect concentration should reach the same minimum for all doping levels at high enough generation rates. 6) From a broader perspective, light-based processing techniques will exhibit the most dramatic effects in semiconductors having compensating defects with shallower charge transition levels, smaller formation enthalpies for the neutral defect to maximize the impact on the formation free energy for charged defects, and large bandgaps, which allow greater modulation of the QFLs. These observations help to explain why light-stimulated epitaxy has produced the most dramatic changes in the free carrier concentrations of doped II-VI and wide bandgap III-V semiconductors. An obvious benefit is that light-assisted processing techniques can be easily implemented. Our model now provides a way to evaluate the outcome of specific processing conditions and optimize them to achieve the greatest reduction in defect concentrations.

While the example of GaSb has been used herein, we reiterate that the model generally treats the behavior of the charge states and capture and emission behavior of the defects, which we assumed to be governed primarily by the charge states. Thus the findings are generally applicable to any semiconductor and any semiconductor deposition or processing technique. This includes the effects of photogenerated carrier populations on compensating defects formed by extrinsic dopants themselves. The model may also explain differences observed in defect concentrations resulting from thermal equilibrium furnace annealing and light-heated techniques such as RTP.

## Acknowledgements

We acknowledge the financial support of the Department of Energy Office of Science, Basic Energy Sciences under contracts DE-AC36-08GO28308 (K.A) and DE-SC0001630 (M.A.S).

## Author Contributions

K.A. and M.A.S. contributed equally to the development of the model, the generation and analysis of the results and the composition and revision of the manuscript.

**Computational Method**

Here we develop the model framework for the simplified case of one extrinsic shallow dopant and one dominant compensating acceptor-like native defect; extension to multiple species is straightforward. We consider a binary compound semiconductor with formula unit AB doped n-type with a fixed density $N_D$ of extrinsic shallow donors and compensated by a fixed density of shallow extrinsic acceptors $N_A$. We assume a thermal equilibrium concentration of a single dominant native defect on the B sublattice ($X_B$) having a dominant charge state, $q$:

$$[X_B^q] = g N_B \exp(-\Delta H_f^q / k_B T) \tag{1}$$

where $g$ is the number of equivalent configurations, $N_B$ is the sublattice site density, and $\Delta H_f^q$ is the defect formation energy. $\Delta H_f^q$ can be calculated as [36]:

$$\Delta H_f^q = E_D + q(\mu_F + E_{VB}) - n_A \mu_A - n_B \mu_B \tag{2}$$

where $E_{VB}$ is the valence band maximum and $E_D$ is the total energy of the supercell comprised of $n_A$ and $n_B$ atoms and a single charged defect. The chemical potentials of $A$ and $B$ are $\mu_A$ and $\mu_B$, and $\mu_F$ is the Fermi level, which represents the chemical potential for both free electrons and holes in thermal equilibrium. The second term in Eq. 2 accounts for the exchange of carriers between the defect and host semiconductor bands and strictly is written for 0 K, where the Fermi-Dirac distribution is a step function. The atomic chemical potentials are restricted to values less than their bulk values, $\mu_{A(bulk)}$ and $\mu_{B(bulk)}$ and the chemical potential of the AB compound is assumed to be the sum of $\mu_A$ and $\mu_B$ such that Eq. 2 can be re-written as:

$$\Delta H_f^q = E_D' + q\mu_F - \frac{1}{2}(n_A - n_B)\Delta\mu \tag{3}$$

in which

$$E_D' = E_D - \frac{1}{2}(n_A + n_B)\mu_{AB} - \frac{1}{2}(n_A - n_B)(\mu_{A(bulk)} - \mu_{B(bulk)}) + qE_{VB} \tag{4}$$

and

$$\Delta\mu = (\mu_A - \mu_B) - (\mu_{A(bulk)} - \mu_{B(bulk)}) \tag{5}$$

The defect's charge transition energies (denoted $q_n/q_m$, e.g. 0/+ for a donor) can be determined by equating the formation enthalpies for the charge states $\Delta H_f^n = \Delta H_f^m$.

Under steady-state excess carrier generation (e.g. photon irradiation), the QFLs $\mu_{Fn}$ and $\mu_{Fp}$ independently describe the concentrations of band electrons and holes. Carriers are exchanged with the extended band states through capture and emission processes rather than thermalization. It is well known that the occupation of a defect state between the bands' QFLs is ambiguously described by the equilibrium Fermi-Dirac functions associated with the QFLs and that detailed balance rate calculations are required to determine the steady-state occupancy.[37,38,39]

Following Simmons and Taylor[39,40], the electron and hole capture rates are:

$$r_{c,n} = v_n \sigma_n N_C \mathbb{F}_{1/2}(\eta_C) N_t (1-f)$$
$$r_{c,p} = v_p \sigma_p N_V \mathbb{F}_{1/2}(\eta_V) N_t f \tag{7a}$$

and the rates of emission are:

$$r_{e,n} = v_n \sigma_n N_C \exp\left(\frac{E_t - E_C}{k_B T}\right) N_t f$$
$$r_{e,p} = v_p \sigma_p N_V \exp\left(\frac{E_V - E_t}{k_B T}\right) N_t (1-f) \tag{7b}$$

in which $v_n$ and $v_p$ are the carrier thermal velocities, $\sigma_n$ and $\sigma_p$ are the capture cross sections, $N_C$ and $N_V$ are the effective densities of states, $\mathbb{F}_{1/2}(\eta)$ is the Fermi-Dirac function, $E_t$ is the trap's charge transition energy, $E_C$ and $E_V$ are the energies of the conduction and valance band edges, $k_B$ is the Boltzmann constant, $T$ is temperature, and $N_t$ is the concentration of traps. The steady-state occupation of a defect by an electron is given by:

$$f = \frac{\bar{n} + e_p}{\bar{n} + e_p + \bar{p} + e_n} \tag{8}$$

wherein

$$\bar{n} = v_n \sigma_n n = v_n \sigma_n N_C \mathbb{F}_{1/2}(\eta_C) \text{ and } \bar{p} = v_p \sigma_p p = v_p \sigma_p N_V \mathbb{F}_{1/2}(\eta_V) \tag{9}$$

$$e_n = v_n \sigma_n N_C \exp\left(\frac{E_t - E_C}{k_B T}\right) \text{ and } e_p = v_p \sigma_p N_V \exp\left(\frac{E_V - E_t}{k_B T}\right) \tag{10}$$

It is not immediately obvious how the formation free energy for a charged defect depends on quasi Fermi energies in the presence of excess carriers since the defect can exchange carriers with both bands. However, interpreting $\Delta H_f^q$ as a statistical average over a large number of identical defect formation events allows an unequivocal result to be derived. As a gedanken experiment, we consider the formation of a population of native defects in the $q=-1$ charge state. This charge state can be achieved starting from the charge-neutral ideal crystal by the formation of a neutral defect combined with 1) the emission of a hole to the valence band reservoir having carrier chemical potential $\mu_{Fp}$, or 2) the capture of an electron from the conduction band having chemical potential $\mu_{Fn}$. The ratio of the transition rates will determine the fractions of the population participating and therefore the weighting that should be given to $\mu_{Fn}$ and $\mu_{Fp}$ in the statistical average.

Thus, in the presence of steady-state excess carriers, $\mu_F$ in Eq. 3 can be replaced for negatively charged states ($q<0$) by:

$$\mu_F = \left(\frac{r_{c,n}}{r_{c,n}+r_{e,p}}\right)\mu_{Fn} + \left(\frac{r_{e,p}}{r_{c,n}+r_{e,p}}\right)\mu_{Fp} \tag{11a}$$

and for positively charged states ($q>0$) by:

$$\mu_F = \left(\frac{r_{e,n}}{r_{e,n}+r_{c,p}}\right)\mu_{Fn} + \left(\frac{r_{c,p}}{r_{e,n}+r_{c,p}}\right)\mu_{Fp} \tag{11b}$$

The value of μF reverts back to the Fermi energy in the case of dark thermal equilibrium.

The computation of the steady-state populations of defects proceeds much as in the equilibrium case by solving for charge balance:

$$\sum_i q[X_i^q] + N_D^+ - N_A^- - n + p = 0 \tag{12}$$

however the coupling of the defect and excess carrier populations via defect-assisted recombination must be accounted for by simultaneously solving the steady-state excess carrier populations:

$$\frac{dn}{dt} = G - U_{BB} - U_{Aug} - \sum_i U_{SRH}\left([X_i^q]\right) = 0$$
$$\frac{dp}{dt} = G - U_{BB} - U_{Aug} - \sum_i U_{SRH}\left([X_i^q]\right) = 0 \tag{13}$$

In Eqs. 12 and 13, *i* indexes over all native defect species, the $N_D^+$ ionized donors ($N_A^-$ ionized acceptors) are determined by $\mu_{Fn}$ ($\mu_{Fp}$), $n = n_o + \Delta n$ ($p = p_o + \Delta p$) are the total carrier densities including the excess electrons $\Delta n$ (holes $\Delta p$), *G* is the steady-state generation rate, $U_{BB}$ is the band-to-band recombination rate, $U_{Aug}$ is the Auger recombination rate, and the $U_{SRH}$ terms account for Shockley-Read-Hall recombination from each of the defects' charge states. The net recombination rates in Eq. 13 are expressed as:

$$U_{BB} = \beta_{BB}\left(np - n_i^2\right)$$

$$U_{Aug} = \left(C_{Aug,n}n + C_{Aug,p}p\right)\left(np - n_i^2\right)$$

$$U_{SRH}\left(\left[X_i^q\right]\right) = \frac{\left(np - n_i^2\right)}{\frac{1}{\sigma_p v_p \left[X_i^q\right]}\left(n + N_C \exp\left[\frac{-(E_{CB} - E_t)}{k_B T}\right]\right) + \frac{1}{\sigma_n v_n \left[X_i^q\right]}\left(p + N_V \exp\left[\frac{-(E_t - E_{VB})}{k_B T}\right]\right)}$$

(14)

in which $\beta_{BB}$ is the band-to-band coefficient, and $C_{Aug,n}$ and $C_{Aug,p}$ are the electron and hole Auger coefficients. All of the quantities in Eqs. 12 and 13 are expressed in terms of $\mu_{Fn}$ and $\mu_{Fp}$, allowing a self-consistent solution to be found for each set of conditions. The model may be extended to account for degenerate carrier densities in the bands.

For the calculations, GaSb was assumed to be either undoped ($N_D = N_A = 0$) or intentionally doped with a shallow donor impurity concentration, $N_D$, of $10^{17}$ cm$^{-3}$ and unintentionally doped with a shallow acceptor impurity concentration, $N_A$, of $10^{13}$ cm$^{-3}$. Defect formation energies were taken from Ref. 11, in which they were calculated using density functional theory with a hybrid functional scheme. Capture cross sections for Ga$_{Sb}$ antisites in the 0, -1, and -2 charge states were assumed to be $\sigma_n$ = 3x10$^{-16}$, 1x10$^{-17}$ cm$^2$, and 5x10$^{-18}$ cm$^2$ and $\sigma_p$ = 3x10$^{-16}$, 1x10$^{-14}$, and 2x10$^{-14}$ cm$^2$, 10$^{-14}$ cm$^2$, ensuring predominant capture of holes, while the capture cross sections for the 0, +1 and +2 charge states for Sb$_{Ga}$ defects were assumed to be $\sigma_n$ = 3x10$^{-16}$, 1x10$^{-14}$, and 2x10$^{-14}$ cm$^2$, cm$^2$ and $\sigma_p$ = 3x10$^{-16}$, 1x10$^{-17}$, and 5x10$^{-18}$ cm$^2$ cm$^2$, ensuring predominant capture of electrons. The capture cross sections of Ga$_i$ defects were assumed to be $\sigma_n$ = 10$^{-14}$ cm$^2$ and $\sigma_p$ = 10$^{-17}$ cm$^2$ respectively. These capture cross-section values are well within the range of those typically accepted in semiconductors, and differ by amounts appropriate for Coulomb attraction(repulsion) for opposite carrier polarities (approximately three orders of magnitude). Variation of these cross-section values did not significantly change the results of the calculation. This can be understood because emission rates depend exponentially on charge transition level but only linearly on cross sections, so only in the rare but commonly-assumed case of recombination centers at $E_i$ with nearly equal cross sections would the model be especially sensitive. Carrier generation rates of $10^{23}$ and $5 \times 10^{24}$ cm$^{-3}$ s$^{-1}$ were examined. Assuming a carrier lifetime on the order of 1 ns, these rates are expected to produce excess carrier concentrations in the range of $10^{14}$ to $5 \times 10^{15}$ cm$^{-3}$.

[39] Simmons, J.G., and Taylor, G.W., Nonequilibrium steady-state statistics and associated effects for insulators and semiconductors containing an arbitrary distribution of traps, *Phys. Rev. B*, **4**, 502 (1971)

[40] Reference 39 contains a number of mistaken formulae.